\documentstyle[epsfig]{mn}

\catcode`\"=\active\let"=\"

\def\3{\ss }

\def\c12{{1\over 2}}

\def\plusplus{\raise 0.3ex\hbox{${\scriptstyle ++}$}{}}
 \def\dedet#1#2{\left(\delta #1\over \delta t\right)_{\rm #2}}

\title[Dynamics of rotating stellar systems -- II]
{Dynamical evolution of rotating stellar systems --
 II. Post-collapse, equal mass system}
\author[E. Kim et al.]
{E. Kim$^{1,2}$, C. Einsel$^2$, H.M. Lee$^{1,3}$, R. Spurzem$^2$, and M.G. Lee$^1$\\
$^1$ Astronomy Program, SEES, Seoul Nat. Univ., Seoul 151-742, Korea\\
$^2$Astronomisches Rechen-Institut, M"onchhofstrasse 12-14,
   69120 Heidelberg, Germany \\
$^3$ Institute of Space and Astronautical Science, Sagamihara, Japan}

\begin{document}

\maketitle

\begin{abstract}

We present the first post core collapse models of initially rotating
star clusters, using the numerical solution of an orbit-averaged 2D
Fokker-Planck equation. Based on the code developed by Einsel \&
Spurzem (1999), we have improved the speed and the stability and included
the steady three-body binary heating source. We
have confirmed that rotating clusters, whether they are in a tidal field or
not, evolve significantly faster than non-rotating ones. Consequences
for observed shapes, density distribution, and kinematic properties of
young and old star clusters are discussed. The results are compared
with gaseous and 1D Fokker-Planck models in the non-rotating case.
\end{abstract}


\section{Introduction}

 Dynamical modeling of globular clusters and other collisional
 stellar systems (like galactic nuclei, rich open clusters,
 rich galaxy clusters) still suffers from severe drawbacks.
 They are due partly to the poor understanding of the
 validity of assumptions
 used in statistical modeling based on the Fokker-Planck and
 other approximations on one hand, and due to statistical
 noise and the impossibility to directly model realistic particle numbers
 with the presently available hardware, on the other hand. For rich
 star clusters with $10^5$ or more particles this statement is still
 true, despite of continuous progress in the development and application of
 special purpose computers (cf. e.g. Makino \& Hut 2001).

 Therefore a detailed comparison of the different methods
 is very important using the same choices of parameters
 and initial data. For spherical, non-rotating star clusters
 this has been provided in the last decade (Giersz \&
 Heggie 1994a,b, henceforth GHI, GHII, Giersz \& Spurzem 1994,
 henceforth GS,
 Spurzem \& Aarseth 1996, Spurzem 1996).
 The basic idea of such comparisons is to check the validity of
 models based on statistical mechanics, e.g. by using the
 Fokker-Planck approximation, against direct orbit integration
 of large $N$-body systems using state-of-the art direct
 $N$-body codes. Depending on the model a number of free
 parameters (such as a free factor in the argument of the
 Coulomb logarithm, to mention one typical example) can
 be adjusted. Since the number of free parameters, however,
 is small, a positive result in the comparison for many
 values of $N$ (particle number) and different initial models
 is considered as a proof of the underlying physical concepts
 and approximations. Such systematic comparisons even across
 a larger number of participating groups have a tradition in
 the field of collisional $N$-body and star cluster modelling
 (Lecar \& Cruz-Gonz\'alez 1972, Heggie et al. 1998) and will
 be continued in the future (Heggie 2001).

 On the side of theoretical or statistical mechanics we mainly
 use the Fokker-Planck equation and the
 direct numerical solution of its orbit average. One example is 
 1D models (assuming that the distribution function $f$ only depends
 on a single constant of motion, the energy $E$ of a stellar
 orbit in the given spherical potential),
 applicable for isotropic spherical systems (Cohn 1980). There are
 generalizations to 2D anisotropic models, where $f$ depends on two
 independent variables, energy and angular momentum $J^2$ (Cohn 1979,
 Takahashi 1995, 1996, 1997, Takahashi, Lee \& Inagaki 1997), thus including
 the possible difference between radial and tangential velocity dispersions
 in a cluster and (in the last cited paper) the effect of a tidal boundary.
 For recent models including tidal fields and a stellar mass spectrum
 compare e.g. Takahashi \& Lee (2000) and Takahashi \& Portegies Zwart (1998, 2000).
 Another type of statistical models is so-called gaseous models, which solve
 numerically, similar to solutions of gas dynamical equations,
 a set of moment equations of the Fokker-Planck equation. They are
 usually taken up to second order and terminated by a heat flux
 equation in third order (Heggie 1984), including as well
 the anisotropy (Louis \& Spurzem 1991, Spurzem 1994).

 On the other side there are direct $N$-body models using
 standard $N$-body codes
 (NBODY5: Aarseth 1985, Spurzem \& Aarseth 1996; NBODY4:
 Makino \& Aarseth 1992, Makino 1996, Aarseth \& Heggie 1998,
 NBODY6 and NBODY6++: Aarseth 1996, 1999a, Spurzem 1999, 
 Spurzem \& Baumgardt 2001) or Monte Carlo
 schemes (Giersz 1996, 1998, 2001, Joshi, Rasio \& Portegies Zwart 2000,
 Joshi, Nave \& Rasio 2001). The Monte Carlo models, being comparable
 in their wealth of data and stochastic nature to the direct $N$-body
 models, still employ the Fokker-Planck approximation.

 Orbit-averaged Fokker-Planck and direct $N$-body models in
 particular have been improved sometimes by including very detailed 
 astrophysical
 ingredients, such as a stellar mass spectrum
 (Spurzem \& Takahashi 1995,
 Giersz \& Heggie 1997, Heggie et al. 1998), adding to that
 a more detailed treatment
 of the tidal truncation (Takahashi, Lee \& Inagaki 1997), or
 approximate modelling of gravitational shocking (Gnedin, Lee \& Ostriker 1999),
 or very detailed treatments of single and binary stellar evolution,
 coupled with a detailed follow-up of merging and collisions
 (Portegies Zwart et al. 1999, Hurley et al. 2001).
 The cited papers are just recent examples and not exhaustive;
 the interested reader could get a better overview e.g. by
 looking at the proceedings volume of Deiters et al. (2001).
 The comparison of different ways of modelling, though not
 as exhaustive yet as in the idealized single mass case,
 has been done at least by studying some selected problems
 regarding tidal mass loss in direct Fokker-Planck versus 
 $N$-body models (Portegies Zwart \& Takahashi 1999, Takahashi
 \& Portegies Zwart 1998, Baumgardt 2001).

 The results could be summarized by
 saying that in general the Fokker-Planck approximation (small
 angle two-body scattering dominates the global evolution of
 the system), the approximation of heat conduction (its energy
 transport can be treated as heat conduction in a collisional
 gas), and the statistical binary treatment (model of energy
 generation by formation and subsequent hardening of three-body binaries
 using simple semi-analytical estimates) all appear to be a
 fairly good description of what happens in $N$-body simulations.
 But there are still two basic drawbacks: 

 (i) all
 comparisons are so far limited to rather small particle
 numbers ($N\le 64000$) as compared to real particle numbers
 of globular clusters of the order of a few $10^5$ or even
 up to $10^6$ stars. Low $N$
 models cannot be easily extrapolated to higher $N$, since
 after core collapse a variety of different processes (close
 encounters, tidal two-body encounters, effects of the finite size
 of the stars) all vary with time scales,
 which depend on different powers of the particle number
 (see e.g. the scaling problem tackled by Aarseth \& Heggie 1998);

 (ii) during core bounce and binary driven post-collapse
 evolution an individual $N$-body simulation exhibits stochastic
 fluctuations, due to the stochastic occurrence of
 superelastic scatterings of very hard binaries with field stars
 and other binaries.
 Although the averaged evolution of the system, understood
 either as a time average (looking for long post-collapse times)
 or as an ensemble average (averaging statistically independent
 single $N$-body models), is reproduced well by the
 theoretical models based on the above assumptions, the
 {\sl individual} evolution of a stellar system, even with
 a relatively large particle number, might not be exactly matched
 at any instant. The collaborative experiment in
 this area (Heggie et al. 1998) gives a good overview:
 all methods do agree fairly well, but variations of quantitative
 results of some 10 or 20 \% and some scaling problems, which
 are not exhaustively examined, have to be tolerated.

 Very few attempts yet have been done to include a rather
 important piece of realism, the existence of an initial
 angular momentum of the star cluster. While the probability
 that the cloud from which a star cluster originates, has
 zero total angular momentum is very small, practically all
 models have assumed that. The inclusion of angular momentum
 requires a much more complex physical treatment, including
 axisymmetry in coordinate space (flattened clusters) and
 a way to deal with a third integral of motion of individual
 stellar orbits, which is not known analytically in general.

 First semi-analytical models by Agekian (1958) assumed that
 escaping stars have the same angular momentum distribution with 
 the remaining ones, and the system keeps a structure as a 
 MacLaurin spheroid. In particular the first approximation is
 rather doubtful, as one could already deduce from an unpublished
 early direct Fokker-Planck model by Goodman (1983). Goodman's work
 was revisited and improved by a much more detailed numerical
 2D Fokker-Planck study of core collapse of rotating star clusters
 by Einsel \& Spurzem (1999, henceforth Paper I, see also Spurzem 2001). 
 In a slightly
 different approach Hachisu (1979) argued that there would be an
 internal, relaxation driven redistribution of angular momentum,
 similar to the mechanism of gravothermal collapse. In Paper I
 it was found that such internal redistribution occurs in the
 early evolutionary phase, but later mass loss and relaxation
 due to angular momentum loss connected with it dominates the
evolution. Also it was found there (by inspection of the numerical
 results), that in the late stages
 of core collapse a self-similar solutions exists in which the
 rotational velocity scales with the same power as the velocity
 dispersion; this finding was underlined by some theoretical
 arguments of Lynden-Bell (2001).

 Paper I demonstrated, that the influence of rotation on star cluster
 evolution is not small. The core collapse time could be accelerated
 very much (that was an equal mass model). So, it is not clear,
 what is the combined influence of rotation and the other processes,
 which had been improved and included into
 cluster dynamics in the past (stellar evolution,
 mass spectrum, tidal fields and tidal shocking, primordial binaries).

 Unfortunately, rotation, though it is a natural initial condition
 from collapse of a star--forming cloud, could not be included in
 most of the existing evolutionary models of star clusters. Monte Carlo
 and Fokker-Planck techniques (with the exception of Paper I) were
 limited to spherical symmetry, as well as gaseous models. A generalization
 of such models poses significant challenges, such as what is the
 effective viscosity scaling describing properly viscous effects due
 to two-body relaxation (Goodman 1983) in the case of gaseous models.
 For Fokker--Planck models the main problem is the requirement to
 neglect a possibly existing third integral of motion on axisymmetric
 potentials, because it cannot be given analytically. In Paper I diffusion
 of orbits was considered disregarding the third integral, i.e. in a
 2D model only considering $E$ and $J_z$, orbital energy and $z$-component
 of angular momentum of a stellar orbit, and a discussion of possible
 errors was given.

 In this paper we continue and improve the work of Paper I in several
 respects. First we include a statistical binary 
 energy generation, to model the heating of the
 core of the cluster in late stages of core collapse due to the
 formation and subsequent superelastic scatterings of hard binaries in
 three-body encounters (Hut 1985, Lee, Fahlman \& Richer 1991). 
 So, the evolution of the cluster, its shape
 and internal parameters, such as density distribution, rotation curve,
 velocity dispersions and mass loss in a steady tidal field, will be
 followed past the time of maximum collapse, where the models of
 Paper I stopped. These are the first post-collapse Fokker-Planck models of
 rotating star clusters published so far.

 Furthermore, we have improved the quality of the numerical integration scheme
 to have a better ground for comparisons with other methods, and an attempt
 has been made to disentangle the effects of rotation and tidal cutoff, which
 was not clear in Paper I, by comparing isolated models (spatially extrapolated
 models to infinity without energy cutoff, but otherwise following the density
 structures of our King models) with the tidally truncated ones.
 In future work $N$-body models will be provided for comparisons as well.
 Some preliminary results can be obtained from Boily (2000) and 
 Boily \& Spurzem (2000). They show that for moderate amounts of rotation
 the Fokker-Planck and $N$-body models agree rather well.
 Also, we compare in the non-rotating case the results of our code with
 standard models (1D Fokker Planck codes by H.M. Lee and K. Takahashi, 
 gaseous model by R. Spurzem).

 It should be noted that our models are still on a rather idealized
 level: there is no stellar mass spectrum, no effects of finite size
 stars, no time dependent tidal field, no primordial binaries. However,
 we stress the importance to study first the undisturbed physical effect
 of rotation on the standard  picture of star cluster evolution. Since
 rotation is such a fundamental initial and natural physical parameter,
 this is considered as a per se interesting study of dynamics of $N$-body
 systems, but it has also strong astrophysical relevance, because the
 characteristics of rotation
 may be a parameter which is less severely influenced than
 others by core collapse and post collapse re-expansion, at least around
 the half-mass radius. So it is a possible diagnostic tool to determine
 dynamical ages of clusters and to set constraints on the cluster's
 initial configurations. Also, galactic nuclei containing supermassive
 black holes are mostly rotating systems, and the physical and
 numerical techniques developed here find their application also for 
 a consistent time-dependent modeling of galactic nuclei.

This paper is organized as follows. In the next section, we briefly
describe our models. In \S 3, we present the results of the
post-collapse cluster evolution of initially rotating clusters, and discuss
the implications of our results.
The conclusion is given in the last section.

\section{The models}

\subsection{Numerical Method}

 A computational scheme to solve the 2D Fokker-Planck (henceforth FP)
 equation with high accuracy for pre- 
 and post-collapse has been worked out. The
 framework of the method is the same as that of Paper I: it is comprised
 of two steps, one is the FP step, in which the distribution function is 
 advanced by solving the FP equation with the gravitational potential
 being held fixed. In the Poisson step, the potential is advanced by
 solving Poisson's equation with the distribution function being
 held fixed as a function of the adiabatic invariants.
Henceforth, the code developed to study dynamical star cluster with 2D 
Fokker-Planck equation is called FOPAX.

 In the following we describe some improvements from previous works.
 First we have improved the
 Poisson step, on which the global accuracy of the computation depends very much,
 as noted in Paper I. We have changed a part where the
volume of the hypersurface for given energy and angular momentum
and the adiabatic invariant are calculated (see eqs. (5) \& (7) in Paper I).
In the previous version of code these quantities are 
calculated by using
simple trapezoidal rule. Now we use a
two dimensional Gaussian quadrature scheme.
A comparison of the adiabatic invariants calculated by using these
two different schemes showed 
 that stars with nearly circular orbits around the cluster center are more
accurately computed with the new scheme. We note also, that
the diffusion coefficients become more accurate than the previous code
thanks to the improved accuracy in computing the volume of the hypersurfaces.

As for the FP step, an essential difference between the method used 
here and that in Cohn (1979) is
concerning the discretization: we apply a finite difference scheme,
where the Chang-Cooper scheme is applied only for the energy direction
(compare also for the anisotropic spherical 2D case by Takahashi 1995). 

In order to extend the evolution beyond the core collapse, we need to 
add an energy source that drives the post core collapse evolution. Primordial
binaries and massive stars can provide energy from very early on, and
they delay significantly the core collapse time and affect the
details of the binary distribution very much (Gao et al. 1991, Giersz
\& Spurzem 2000). In order to compare well our models with previous standard
results, and due to a lack of any good method to include many hard binaries
in our model, we have only considered the heating effect due to
three-body binaries.

The energy generation rate by three-body binaries per mass unit is given as
(e.g., Hut 1985)

\begin{equation}
\dedet{e}{\rm 3b} = C_b {\rho^2\over m^2\sigma^2}
  \Bigl({Gm\over\sigma}\Bigr)^5 \ \ .
\label{eq-edot}
\end{equation}
Here $\rho$ and $\sigma$ are the local mass density and 1D velocity
dispersion, respectively,
$G$ the gravitational constant and $m$ the individual stellar
mass.
It is shown in Giersz \& Spurzem (1994)
and Giersz \& Heggie (1994a,b)
that for particle numbers between $N=1000$ and $N=10000$ the
best agreement between direct $N$-body calculations, direct solutions
of the orbit-averaged Fokker-Planck equation and anisotropic
gaseous models is achieved for one set of parameters, including
the parameter of the Coulomb logarithm $\gamma = 0.11 $, and
$C_b=90$. The latter value used to be the standard value derived
from theoretical arguments, based on a numerical factor of
${\tilde C}=0.9$ in the formula for the formation rate of
three-body binaries (Hut 1985). It has been argued (Goodman \& Hut 1993),
that ${\tilde C}=0.75$ is a better choice, but still within
some uncertainty. The results of comparisons with $N$-body simulations
show that $C_b=90$ is a fairly reasonable value, but within the
uncertainty $C_b=75$ (which would ensue with the new formation
rate) cannot be ruled out. However, we also note that
the detailed evolution is very
insensitive to this constant.

\subsection{Initial Models and Boundary Conditions}

As in Paper I, we employ the rotating King models as initial
models following Lupton \& Gunn(1987).
These models are characterized by two parameters:
dimensionless central potential $W_0$ and the rotational
parameter $\omega_0$. We have examined the
evolution of clusters with $W_0=6$ and $W_0$=3. The rotational
parameters are chosen such that the cluster remains to be stable
against the dynamical instabilities. In Table 1, we have
listed the global parameters of the initial models used in the
present study. The rotation parameters for models with central potential
$W_0 = 3$ are chosen such that the ratios of initial rotational energy to
initial potential energy should be similar to those for models with
central potential $W_0 = 6$ as shown in Table 1.
For uniformly rotating
systems, secular instability is known to arise if $T_{rot}/
|W|> 0.14$ (Ostriker \& Peebles, 1973), where $T_{rot}$ is the 
rotational kinetic energy and $W$ is the potential energy. 
All of our models satisfy the stability criterion.

\begin{table}
\centering
\begin{minipage}{85mm}
\caption{Properties of Initial Models}
\begin{tabular}{@{}crrrcccc@{}} \hline\hline
$W_0$ & $\omega_0$ & $r_t/r_c$ & $r_h/r_t$ & 
$T_{rot}/|W|$ & $e_{dyn}$ & $r_h$& $\tau_{rh}$ \\
      &            &           &           &
              &           & [pc] & [yrs] \\[3pt] \hline\hline
  & 0.0 & 18.0 & 0.15 & 0.000 & 0.000 & 4.19& $1.64\times 10^8$ \\
6 & 0.3 & 14.5 & 0.18 & 0.035 & 0.107 & 5.02& $2.15\times 10^8$ \\
  & 0.6 &  9.6 & 0.24 & 0.101 & 0.285 & 6.70& $3.32\times 10^8$ \\[1pt] \hline
  & 0.0 &  4.7 & 0.26 & 0.000 & 0.000 & 7.25& $3.73\times 10^8$ \\
3 & 0.8 &  4.2 & 0.29 & 0.035 & 0.102 & 8.89 & $4.40\times 10^8$\\
  & 1.5 &  3.3 & 0.35 & 0.097 & 0.267 & 9.77& 5.84$\times 10^8$ \\ \hline
\end{tabular}

$e_{dyn}$: dynamical ellipticity as defined in Paper I.\\
$T_{rot}/|W|$: rotational over potential energy.\\
$m=1 M_\odot$ assumed to compute dimensional quantities.
We have used fixed value of $r_t=27.9$ pc to obtain $r_h$.

\end{minipage}

\label{tab-inimod}
\end{table}

\begin{figure}
\epsfig{figure=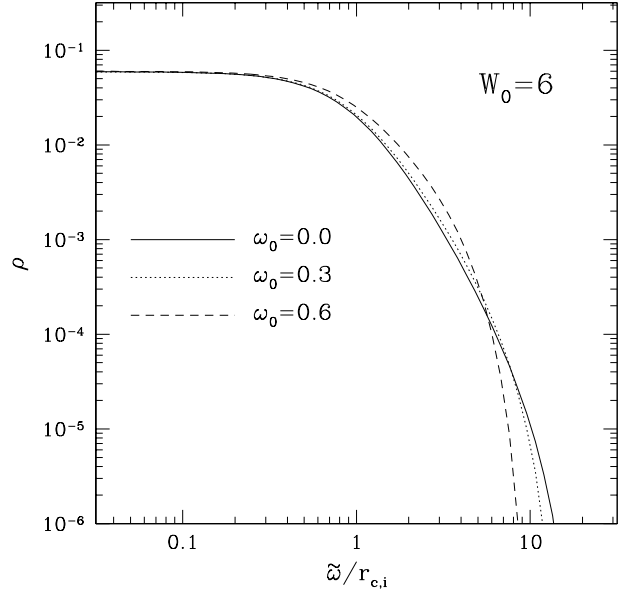, height=0.475\textwidth, width=0.475\textwidth}
\vspace{-3mm}
\caption{The density profiles along the major axis of selected
initial models. Rapidly rotating model is less centrally concentrated
than slowly rotating or non-rotating model.
}
\label{fig-ini_den}
\end{figure}

The density structures along major axis
of the initial models for $W_0=6$
and $\omega_0=0$, 0.3 and 0.6 are shown in Fig. \ref{fig-ini_den}.
The density profiles are similar to each other except near the
tidal boundary. As we have seen from Table 1, the $r_t/r_c$ becomes
smaller as $\omega_0$ increases. Therefore, the
rapidly rotating clusters are less centrally concentrated than 
slowly rotating or non-rotating clusters with the same $W_0$. 


We have considered two different boundary
conditions: tidally limited and isolated cases. For the tidally
limited models, we assumed that the cluster orbits in
the spherical potential of its mother
galaxy at a constant distance (circular orbit),
so that the mean density within the tidal radius ($r_t$) remains 
a constant throughout
the evolution. We have removed stars beyond the tidal radius
instantaneously at every Poisson step. This boundary condition
is clearly somewhat extreme. In actual clusters, the stars remain
near the tidal boundary for orbital time scales. Depending
on the treatment of the evaporation process, the lifetime of
the clusters could vary significantly, especially for small
$N$ clusters (Takahashi \& Portegies Zwart 2000). As we will specify 
later, our models actually use rather small $N$ in order to
reduce the computing time (see below). By using small $N$, the core
collapse stops at relatively small core density. However, the
outer parts are not sensitive to the choice of $N$,
except for evaporation times (in units of half-mass relaxation
time). We can interpret our results as those of large $N$ for
the long-term evolution of the global properties.

For the case of the isolated models, we have to extend the calculation to
infinite distance which is impossible. Thus we set the computational
boundary at $10\,r_{t,i}$, where $r_{t,i}$ is the tidal radius of the
initial model. If the stars reach beyond
this radius, they are `removed'. The mass lost by this process is usually
very small. A similar boundary condition was used in computing the evolution
of isolated clusters in previous studies (see, for example, Quinlan 1996). 
We found that very little mass is lost for isolated clusters until the
end of computation with this boundary condition. Therefore we can
study the influence of rotation on relaxation alone, without interference
of the (tidal) mass loss processes.

The pre-collapse evolution does not depend on the total number of stars
($N$). However, the relative importance of the binary heating to the
two-body relaxation depends on $N$. Larger $N$ means higher 
central density at the time of expansion, and thus longer computational time.
If $N$ is too small, we cannot apply the Fokker-Planck method. 
As a compromise, we
have used $N = 5000$ for the models considered in the present paper.
This is obviously much smaller than that of  actual globular clusters, 
and brings relaxation time and crossing time closer to each other than
in real clusters. According to
recent comparisons between direct FP and $N$-body models (Takahashi \&
 Portegies Zwart 1998, Portegies Zwart \& Takahashi 1999) the evolution of
the total mass of the cluster until total dissolution depends rather 
sensitively on
the particle number, because the mass loss process involves the crossing
time rather than the relaxation time. Nevertheless we restrict ourselves here
 to models using relatively small $N$, because this speeds up the computation
 (not so deep central collapse is reached) and we mainly want to study here
 the interplay between rotation, relaxation and tidal mass loss, where it is
 an advantage to have the relevant time scales not too different from each other.
Also, $N = 5000$ is already large enough to study the effect of core bounce caused by
formation and hardening of three-body binaries. A more detailed and
realistic modelling should include more cluster physics anyway, so
we do not provide models with large $N$ here, but restrict ourselves to this rather
theoretical study of relaxing rotating collisional star clusters.

\section{Results and Discussions}

\subsection{Non-rotating Limit}

\begin{figure}
\epsfig{figure=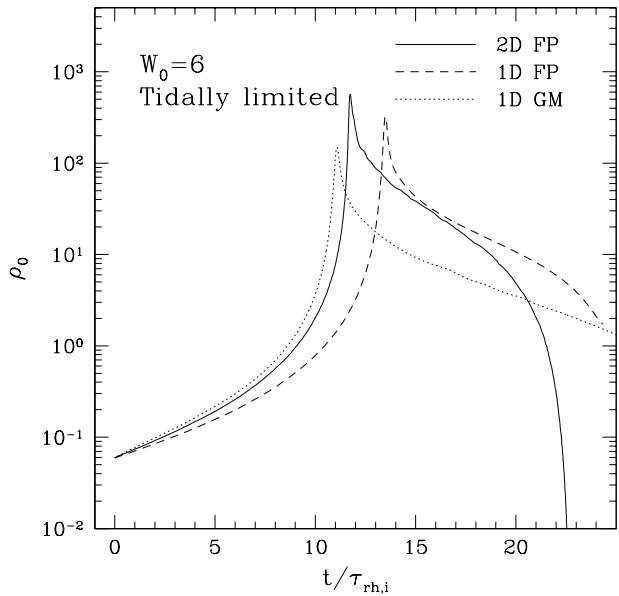, height=0.475\textwidth, width=0.475\textwidth}
\vspace{-3mm}
\caption{Comparison of central density evolution for model with ($W_0=6, \omega_0=0$) between FOPAX
(solid line), 1D FP (dashed line) and 1D gaseous
model (dotted line).}
\label{fig-cden}
\end{figure}

First, we compare
the evolution of tidally limited clusters by FOPAX 
(assuming $\omega_0=0$), 1D FP and gaseous models.
The evolution of central density of tidally limited, 
non-rotating models with 
the central King potential of $W_0=6$ is shown 
in Fig. \ref{fig-cden}.
The results obtained by the 1D FP code
(Lee, Fahlman \& Richer 1991) and 1D gas model (e.g., Spurzem 1996), are shown 
as broken and dotted lines, respectively.
The time is expressed in units of initial half-mass relaxation time
($\tau_{rh,i}$), and the central density in units of $M_i/r_{c,i}^3$, where
$M_i$ and $r_{c,i}$ are initial mass and King's core radius, respectively.

There are some differences in collapse times among different
methods. 1D Fokker-Planck model has the longest and the 
gaseous model has the shortest collapse times. Quinlan (1996)
reported that the isotropic 1D model with initial $W_0=6$
reaches the core-collapse at 12.94 $t_{rh,0}$, which is about 6\% shorter
than our 1D model shown in Fig. \ref{fig-cden}. Part of the difference
is due to the binary heating: when the binary heating is turned off,
the core collapse time becomes around 13.4 $t_{rh,0}$, which is
about 3.5\% longer than Quinlan (1996), but in perfect agreement 
with another type of calculations by K. Takahashi
(private communication) using the anisotropic Fokker-Planck code in
$(E,J)$ where development of anisotropy is suppressed. 

The FOPAX, applied to an initially non-rotating
cluster reaches the core collapse in shorter time than any of the 
other isotropic 1D models. 
The central density for 2D FP model at the time of core collapse is
larger by a factor $\sim 2$ than that of 1D FP model.
In principle, these calculations should give
the same result, but because of different treatment of the detailed
numerical integration, we regard these discrepancies as insignificant
for the general behavior of the long term evolution. The lifetime
of the rotating 2D Fokker-Planck model is also shorter by about
10 \% than isotropic model.

The collapse time for 1D gas model is 
shorter than other computations by about $5 \sim 20\%$. 
The difference of collapse
time between gas model and FP model is also seen elsewhere
(Spurzem \& Takahashi 2001), where it is interpreted as a result
of different physical approximations used in gas and Fokker-Planck models.

\subsection{Isolated and Tidally Limited Models}

\begin{figure}
\epsfig{figure=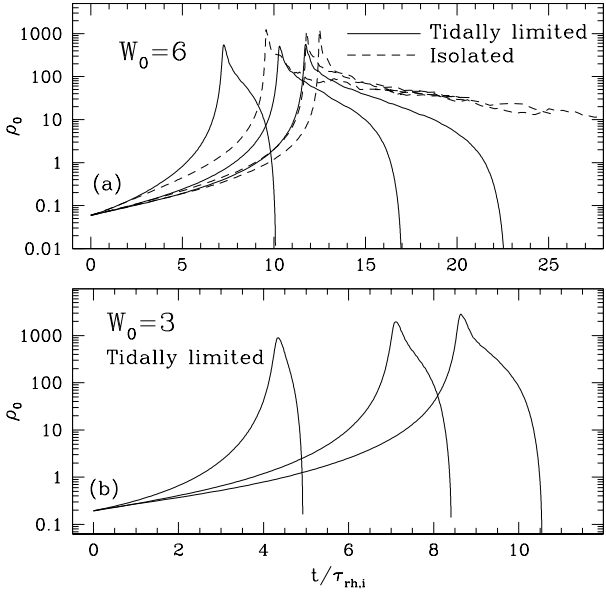, height=0.475\textwidth, width=0.475\textwidth}
\vspace{-3mm}
\caption{Comparison of central density evolution for models with $W_0=6$ (a),
         and for models with $W_0=3$ (b). The initial degree of rotations are
         $\omega_0 = 0.6, 0.3, 0.0$  for $W_0=6$ and
$\omega_0=1.5, 0.8, 0.0$ for $W_0$=3, from left to right. The evolution of
         central density for models without tidal boundaries also shown
         (dashed lines in (a)).}
\label{fig-cden2}
\end{figure}

Fig. \ref{fig-cden2} displays the time evolution of central density 
for our 9 models.
Solid lines represent the run of central density with tidal boundary and
dashed lines for isolated cases. The acceleration of core
collapse due to rotation is shown clearly for both types 
of boundary conditions.
The tidally limited rotating models reach core collapse earlier 
than non-rotating ones, when the time is measured in units of
half-mass relaxation time $\tau_{rh,i}$, as reported in Paper I. 
However, the evolution of
central density of isolated stellar systems was not computed
previously. 
The acceleration of core collapse owing to initial rotation 
for an isolated stellar system is clearly visible
in Fig. \ref{fig-cden2}a. However, the acceleration of core
collapse in a rotating star cluster (as compared to the non-rotating models) is
stronger if also tidal mass loss is present (which was the case only studied in
Paper I). So we have distinguished the two major effects accelerating core
collapse in rotating star clusters, one is due to enhanced two-body relaxation
alone, and the other is a coupling of rotation with mass loss.

\begin{table}
\centering
\begin{minipage}{95mm}
\caption{Time-scales of tidally limited cluster models.}
\begin{tabular}{@{}crrrrcc@{}} \hline\hline
$W_0$ & $\omega_0$ & $t_{cc}/\tau_{rh,i}$ & $t_{dis}/\tau_{rh,i}$ & 
${{t_{dis}-t_{cc}} \over {M_{cc} /
\sqrt{G \bar{\rho}}}}$ & $M_{cc}$ & $t_{50}/\tau_{rh,i}$ 
\\[3pt] \hline\hline  
  & 0.0 & 11.73 & 22.61 & 0.119 &0.59 & 13.2 \\
6 & 0.3 & 10.31 & 16.96 & 0.123 &0.48 & 10.1\\
  & 0.6 &  7.27 & 10.08 & 0.144 &0.33 & 5.4 \\[1pt] \hline
  & 0.0 &  8.63 & 10.54 & 0.375 & 0.25& 5.4\\
3 & 0.8 &  7.12 &  8.41 & 0.405 &0.18 & 4.0 \\
  & 1.5 &  4.35 &  4.92 & 0.499 &0.09 & 1.7 \\ \hline
\end{tabular}

$t_{cc}$: core collapse time.\\
$t_{dis}$: dissolution time of clusters.\\
$t_{50}$: time at which the cluster lose half its total mass.\\
$M_{cc}$: represents the mass retained in a cluster at $t \approx t_{cc}$.

\end{minipage}
\label{tab-tscales}
\end{table}

As seen from Fig. \ref{fig-ini_den}, rotating clusters tend to
have smaller concentration: for models with
$W_0=6$, $r_t/r_c$ decreases from 18 for $\omega_0=0$
to 9.6 for $\omega_0=0.6$. It is well known that 
the time to core-collapse in units of $\tau_{rh,i}$
decreases with increasing $r_t/r_c$ for isolated models, but we find the
opposite trend for rotating models. Clearly, rotation plays an
important role in accelerating the core-collapse.
 
In Fig. \ref{fig-tccnr}, we display the time to core-collapse,
in units of initial relaxation times, as a function of initial
central concentration for non-rotating King models
($W_0=3$ or $W_0=6$, equivalent to a wide range of
initial central concentration $c=\log\left(r_t/r_c\right)$).
In particular we examine here the difference between isolated and tidally limited models. 
For isolated models, $t_{cc}/\tau_{rh,i}$ decreases monotonically with
increasing $c$. This behaviour can be compared with the core-collapse
times of isolated clusters with fixed $W_0 =6$ but different
$\omega_0$.
Contrary to the fact that the non-rotating
isolated models have longer $t_{cc}/\tau_{rh,i}$ for
less concentrated initial models, the rotating models show an
opposite trend. The rotation obviously is fully responsible
for the acceleration of the isolated models.

For the tidally limited models, the behaviour of $t_{cc}$ as 
a function of initial $c$ is not monotonic anymore because of
the role of mass loss. Initially less concentrated models lose
a significant fraction of their mass by the time of core collapse, and thus
the $t_{cc}/\tau_{rh,i}$ becomes smaller for for smaller $c$ starting at
around $c\approx 1.1$ (just below $W_0=6$). Therefore,
we can understand the further reduction of the core-collapse
times for tidally limited models as the action of the mass loss.

The post-collapse evolution of {\em isolated}
rotating and non-rotating systems becomes very similar
after about $t/\tau_{rh,i} > 15$
(dashed curves in upper panel of Fig. \ref{fig-cden2}).
The similarity of post collapse evolution of initially different
models at later times can be understood in terms of simple
energy balance arguments. Goodman (1987) and Lee (1987) showed that
the central density during the post-collapse evolution follows
$(t-t_{cc})^{-2}$ if the major driving energy source is
three-body binaries. The amplitude is mainly determined by the
density at maximum collapse, which is nearly the same for different
$\omega_0$ for a fixed $W_0$ in the models shown in Fig. \ref{fig-cden2}.
At later times the difference in $t_{cc}$ does not play an important
role, and the evolution becomes similar to each other.
The core-collapse can be accelerated by the rotation, but the
post-collapse evolution, which is mostly determined by the
energy balance between the central heating and the overall
expansion, is not sensitive to the presence of the rotation for
the {\em isolated clusters}.

In the presence of a tidal cutoff, however, the simple energy argument
does not provide power law behaviours of
cluster parameters (Lee, Fahlman \& Richer 1991) because
of the mass loss. The rate of mass loss becomes an important factor
in determining the course of post-collapse evolution.
Lee \& Ostriker (1987) showed that the lifetime of the
clusters in steady tidal field is approximately
proportional to $N /\sqrt{G\bar \rho}$, where $N$ is the
number of stars and $\bar \rho$ is the mean density,
which remains constant throughout the evolution if the
tidal field does not depend on time,
of the cluster.

\begin{figure}
\epsfig{figure=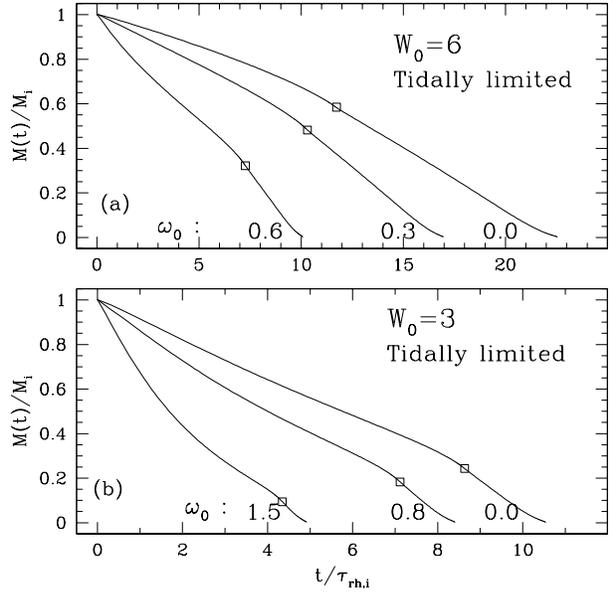, height=0.475\textwidth, width=0.475\textwidth}
\vspace{-3mm}
\caption{Evolution of total mass retained in the cluster models with
         $W_0=6$ (a), and with $W_0=3$ (b). Open squares denote the
         times for core collapse.}
\label{fig-mass}
\end{figure}

We have summarized some important times in units of 
$\tau_{rh,i}$, as well as the time until the complete disruption
from the core collapse, and remaining mass at the time of
core collapse in Table 2. The time evolution of the total
mass of our models is shown in Fig. \ref{fig-mass}.

\begin{figure}
\epsfig{figure=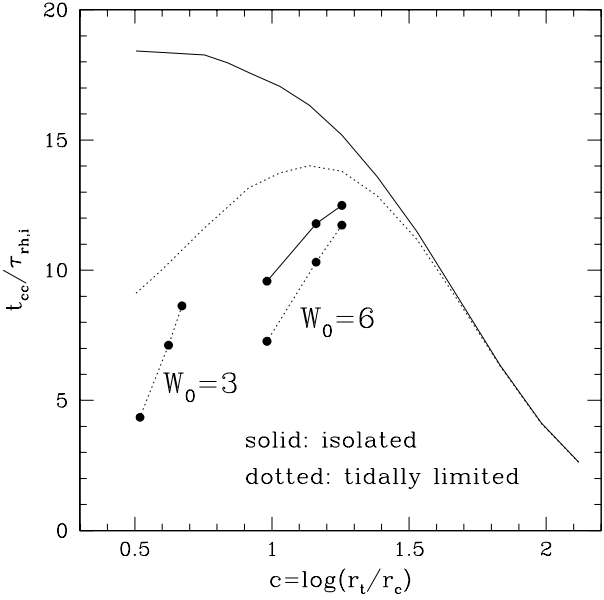, height=0.475\textwidth, width=0.475\textwidth}
\vspace{-3mm}
\caption{The time to core collapse for non-rotating King models as 
a function of the initial central concentration $c(=\log\left(r_t/r_c\right))$.
The core collapse time behaves monotonously with 
$c$ for isolated models, but not for tidally limited models. We also have
shown the collapse times of our rotating models. Note that there is
some discrepancy between the results obtained by 1D FP code and
FOPAX applied to non-rotating clusters, as seen from Fig. 2.}
\label{fig-tccnr}
\end{figure}

\subsection{Evaporation Rates}

Since $\tau_{rh}\propto \sqrt{N} r_h^{3/2}$ and 
$t_{dis}\propto N r_t^{3/2}$, disruption time in units of
$\tau_{rh,i}$ is proportional to $\left(r_t/r_h\right)^{3/2}$.
From Table 1, we
expect that $t_{dis}/\tau_{rh,i}$ would be reduced by about a factor of 2
from $\omega=0.6$ to $\omega_0=0$, for $W_0=6$ models, and by
a factor of 1.56 from $\omega_0=1.5$ to $\omega_0=0$ for
$W_0=3$ models. The disruption time scales listed in Table 2 
for models with different $W_0$ and $\omega_0$ show
somewhat faster evolution of rotating models compared to
that of the scaling law above. This may be interpreted as the
role of the rotation on the mass loss rate.

The evaporation process can be simply understood as a continuous
generation of stars with velocity greater than the escape velocity
($v_e$) through the dynamical relaxation. For isolated uniform stellar
systems, $\langle v_e^2 \rangle = 4\langle v^2 \rangle =
4 v_{rms}^2$ (Ambartsumian 1938;
Spitzer 1940), where the brackets denote the mass-weighted
average over the cluster. If we assume that an equilibrium
velocity distribution $f(v)$ is established in half-mass relaxation time and 
that the stars are removed uniformly, the dimensionless mass
loss rate becomes

\begin{equation}
\xi_e \equiv -{t_{rh}\over M} {dM \over dt} = {\int_{\langle v_e^2\rangle^{1/2}}^\infty
f(v) d^3{\bf v}\over \int_0^\infty f(v) d^3{\bf v}}.
\end{equation}
By employing $\langle v_e^2\rangle^{1/2}  = 2 v_{\rm rms}$ and assuming 
Maxwellian velocity distribution, Ambartsumian (1938) and
Spitzer (1940) obtained $\xi_e = 0.00738$.
For tidally limited models, the escape energy is reduced by
$GM/r_t$ from the simple estimate given above since the
stars are evaporated when they have enough energy to leave tidal 
boundary. By assuming $W = - 0.4 GM^2 / r_h$, the escape
velocity can be expressed as (Spitzer 1987, \S3.2)
\begin{equation}
\langle v_e^2\rangle^{1/2} = 2\left( 1- \lambda\right)^{1/2}~\langle v^2\rangle^{1/2},
\end{equation}
where 
\begin{equation}
\lambda= {GM\over r_t}\ /\ {0.8GM\over r_h} = {5 r_h\over 4r_t}.
\end{equation}
We denote the evaporation rate obtained by using reduced escape velocity as
Ambartsumian-Spitzer (abbreviated as AS)  rate ($\xi_{\rm AS}$).

\begin{figure}
\epsfig{figure=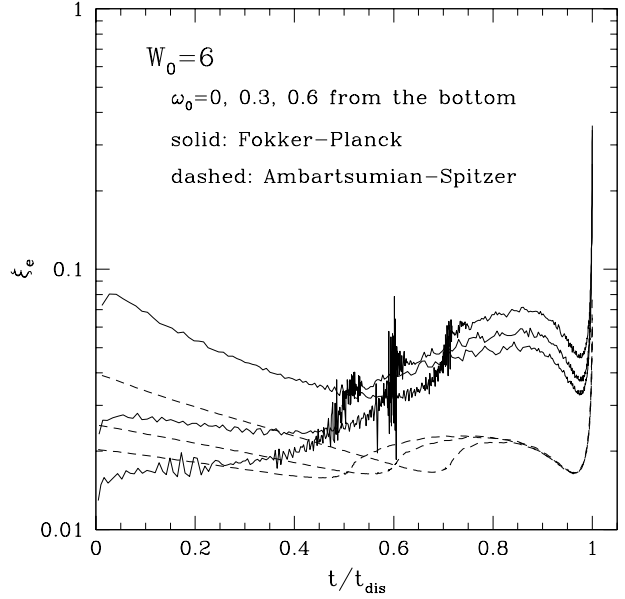, height=0.475\textwidth, width=0.475\textwidth}
\caption{The mass ejection rate computed from our Fokker-Planck models
(solid line) and from Ambartsumian-Spitzer formula, for $W_0=6$
models (dashed line). The noise in estimates of $\xi_e$ near
the collapse is mainly due to the fact that we have kept only
three decimal points in mass estimates while the time step 
becomes very small during these phases.}
\label{fig-xi}
\end{figure}

In Fig. \ref{fig-xi}, we have shown the behavior of $\xi_e$ of our
models with $W_0=6$ together with $\xi_{\rm AS}$ using $r_h/r_t$ data
from the evolving models. The actual $\xi_e$ is usually larger than
$\xi_{\rm AS}$ by a factor of 2 or more. The Ambartsumian-Spitzer formula
assumes that the stellar evaporation takes uniformly over
the cluster. This is clearly a very simplified assumption, and
could be the reason for the discrepancy. Such discrepancies are 
also noticed by Takahashi \& Lee (2000). We simply note that the
general tendency is similar between the model and AS formula. 
As we will see below, large fraction of angular momentum is
removed by the time of post-collapse phase. However, the effect
of rotation seems to be persistent on the stellar evaporation
rate, although the effect is
much smaller than during the pre-collapse. 
During the post-collapse phase, there exists differences  among $\xi_e$ 
with different initial rotation parameters. This might be an indication of the 
importance of the rotation on the stellar evaporation. Although the 
angular momentum disappears rather quickly as the cluster loses
mass, the location of the maximum rotation moves outward so that the
rotation plays some role in mass evaporation. (see below)

%

\subsection{Angular Momentum Transport}

\begin{figure}
\epsfig{figure=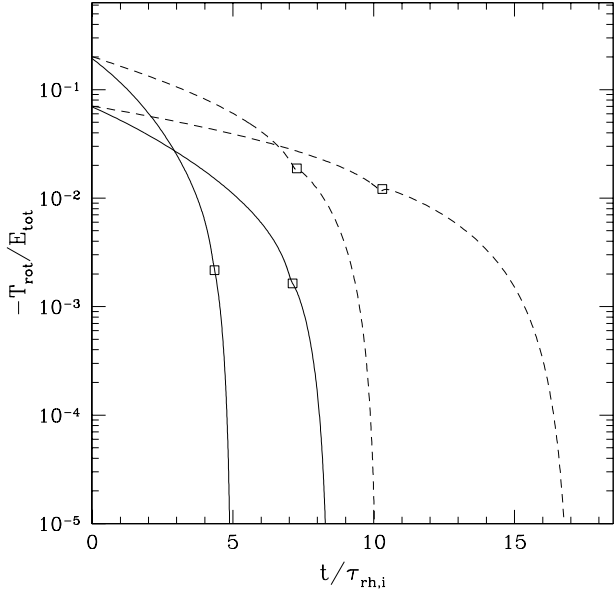, height=0.475\textwidth, width=0.475\textwidth}
\vspace{-3mm}
\caption{Evolutions of rotational energy in units of total
cluster energy for $W_0=3$ (solid lines) and $W_0=6$(dashed lines).
The epochs of maximum collapse are indicated as open squares.}
\label{fig-roten}
\end{figure}

Two-body relaxation causes the angular momentum to diffuse
outward. The stars with high angular momentum preferentially move
outward and eventually escape from the cluster. Therefore, the
angular momentum decreases with time. 
In Fig. \ref{fig-roten}, we have displayed the evolution of
rotational energy in units of total energy for our models.
The epochs of core collapse are indicated as squares in the
figure.
By the time of core collapse, the rotational energy becomes only a very
small fraction of total energy, and the loss of rotational energy is
even accelerated during the post-collapse phase.


Evolution of the rotational speed along the equator can be followed in several
ways. Fig. {\ref{fig-angsp}(a) displays the time evolution of rotational 
speed at half-mass and core radii of the initial models
in equator for models with tidal boundary.
The rotation speed at initial half-mass radii for both models with
different initial degrees of rotation ($\omega_0$= 0.3, 0.6)
decrease monotonically with time. However, evolution of
rotation speed at a radius of initial core radius shows a different structure,
especially for the model with initial degree of rotation of $\omega_0 = 0.6$.
The rotation speed at a radius of initial core radius shows a slight increase
during $t/\tau_{rh,i} \le 2$. Inside the core the rotation structure can
be approximated as a rigid body. Hachisu (1979) claimed that a stellar
system with constant angular velocity should experience gravo-gyro catastrophe
owing to a negative specific moment of inertia. Near the initial
core radius, the rotation speed increases due to the negative specific
moment of inertia.

Some amount of rotation is still present after core bounce as shown in
Fig. \ref{fig-angsp}(b), which displays evolutions of 
maximum rotation speed $V_{rot,{\rm max}}$
in the equator. The maximum rotation speed decreases monotonically throughout
entire evolutionary stage. The rotation speed
decreases rather rapidly after the core collapse. 
The ratio between half-mass radius and the radius in equator
where the rotation speed has a maximum value is shown in Fig. 
\ref{fig-angsp}(c) for tidally limited
models. The model with strong rotation has a value
around $\sim 1.4$ until $t/\tau_{rh,i} \le 3$, then increases gradually
until the collapse. 
The global value of this ratio is around $2$ irrespective of initial
degrees of rotation.

\begin{figure}
\epsfig{figure=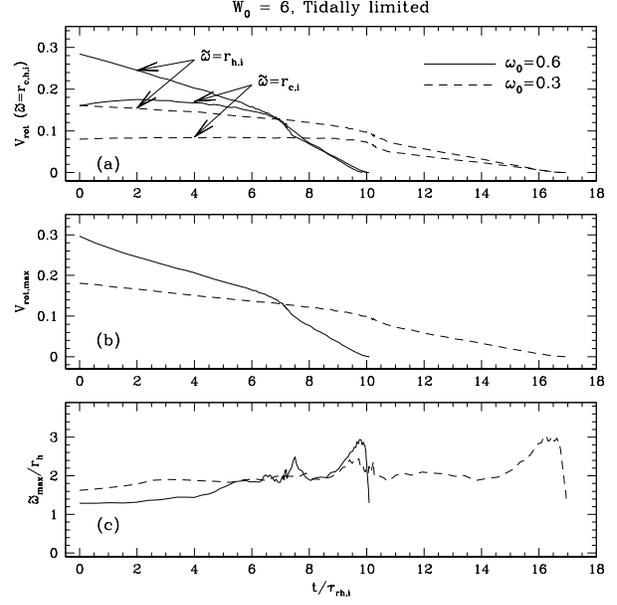, height=0.475\textwidth, width=0.475\textwidth}
\vspace{-3mm}
\caption{Evolutions of rotational velocity for models with $W_0=6$ and
         $\omega_0=0.6$(solid lines) and $\omega_0=0.3$(dashed lines).
         The time evolution of rotational velocity for two fixed radii of
         the initial core and half-mass radii is shown (a). Mid panel (b) shows
         time evolution of the maximum value of rotational velocity at the equator.
         The ratio between the radius where the rotational velocity is 
         maximum in equator and the half-mass radius is displayed in lower panel (c).}
\label{fig-angsp}
\end{figure}

\begin{figure}
\epsfig{figure=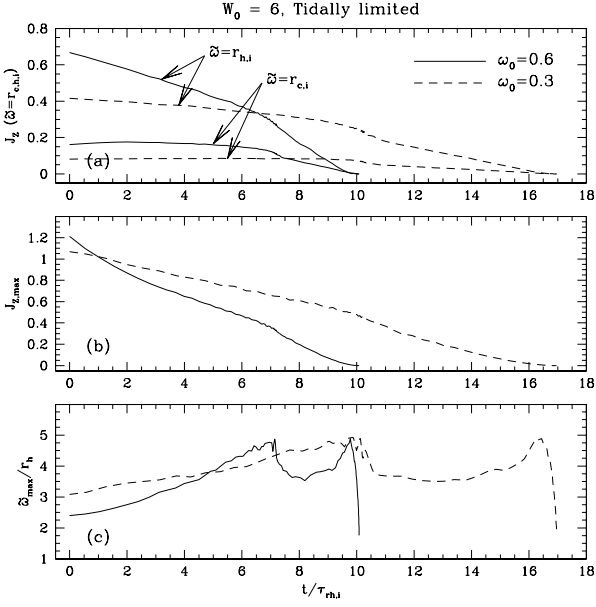, height=0.475\textwidth, width=0.475\textwidth}
\caption{Same as Fig. \ref{fig-angsp}, but for $z$-component of angular momentum
         (see equation (5) for definition).}
\label{fig-spam}
\end{figure}

\begin{figure}
\epsfig{figure=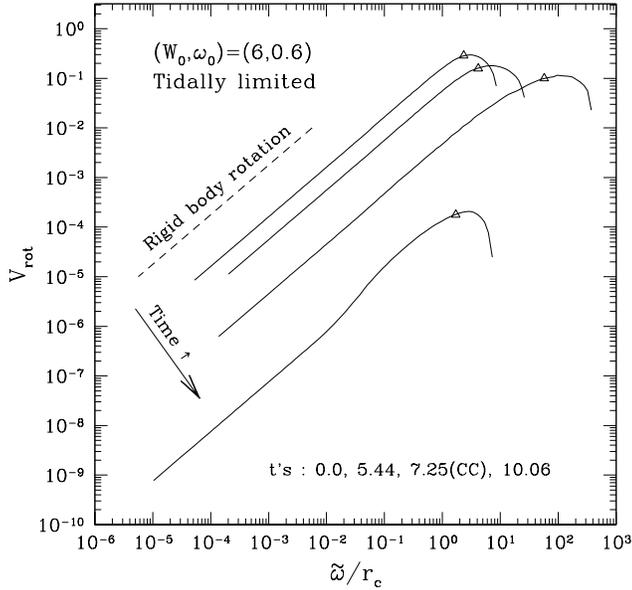, height=0.475\textwidth, width=0.475\textwidth}
\vspace{-3mm}
\caption{Radial profiles of rotational velocity in equator
        for tidally limited cluster model $(W_0,\omega_0) = (6,0.6)$
        at four different evolutionary stages.  
        For comparison, the rotation profile for solid body is shown
        (dashed line). The mass retained in a cluster is about a half of
        initial mass at $t/\tau_{rh,i} = 5.44$. At $t/\tau_{rh,i} = 10.06$
        the cluster is dissolved. Half-mass radii are marked with
        open triangles. Note that the radius is measured in unit of current
        core radius.}
\label{fig-rotcv}
\end{figure}

Fig. \ref{fig-spam} shows the time evolution of $z$-component of 
specific angular momentum which is defined by,

\begin{equation}
J_z(\tilde{\omega}) = v_{\rm rot}(\tilde{\omega}) \tilde{\omega}.
\label{eq-jz}
\end{equation}
Fig. \ref{fig-spam}(a) shows that the angular momentum at $r=r_{h,i}$
decreases faster than the $J_z$ at $r=r_{c,i}$,
where $r_{h,i}$ and $r_{c,i}$ are initial half-mass and core radii, 
respectively, with initial rotation parameter $\omega_0 = 0.6$. A similar pattern
is shown for the model with slower initial rotation, but not clearly. Fig. 
\ref{fig-spam}(b)
displays the time evolution of maximum angular momentum in the equator. 
The location of $J_{z,max}$ in units of $r_h$ is shown in the lower
panel of Fig. \ref{fig-spam}. It clearly shows the outward movement of
angular momentum. 




The rotation curves at selected epochs are shown in Fig. \ref{fig-rotcv}
for $(W_0, \omega_0)=(6,0.6)$. Because of angular momentum loss, the
rotation speed tends to decrease with time, but the shape of the
curve remains nearly the same. The angular speed is nearly
constant (i.e., rigid body rotation) up to $r_h$, and drops rapidly
toward the tidal radius.

\begin{figure}
\hspace{-5truecm}
\epsfig{figure=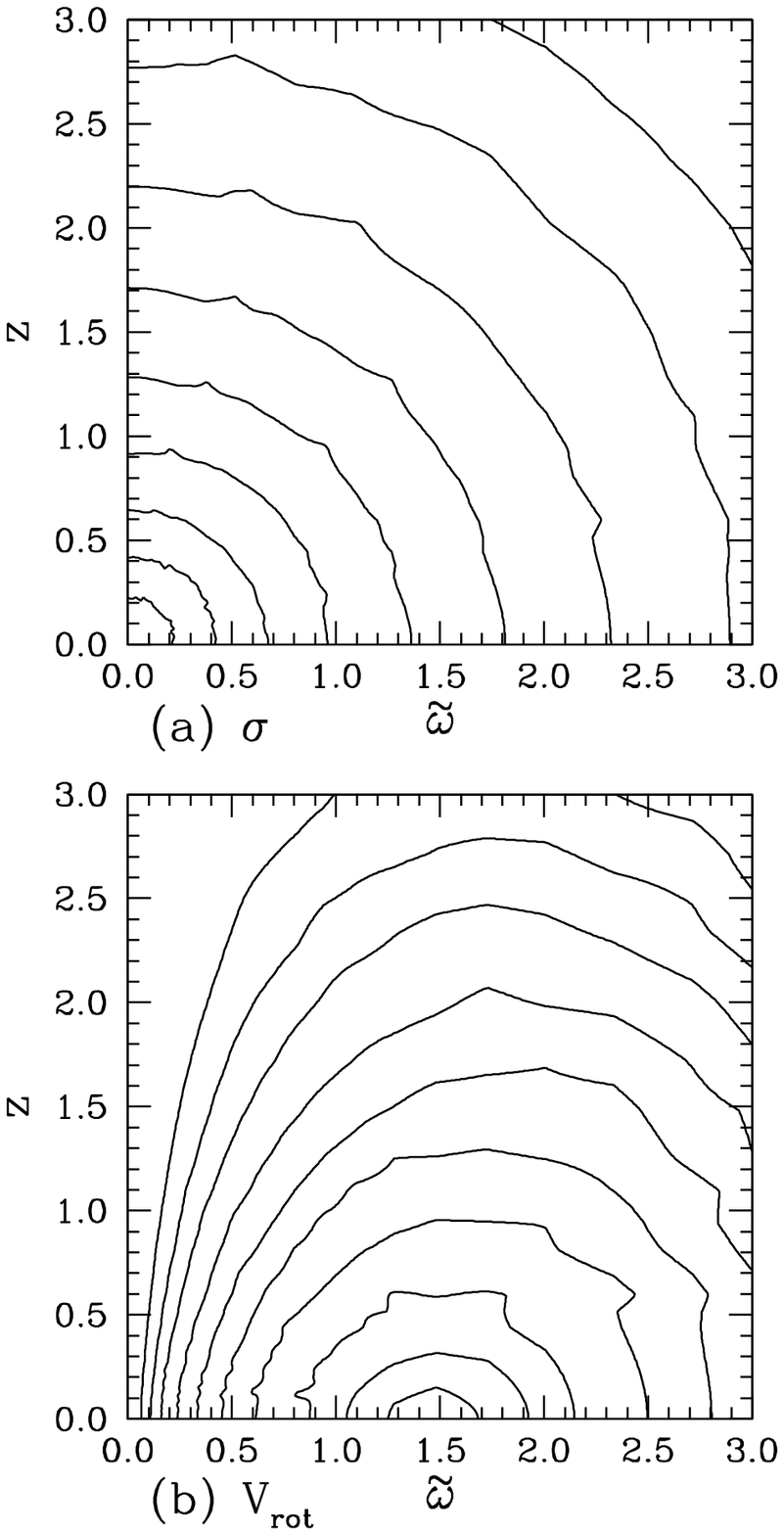, height=0.900\textwidth, width=0.900\textwidth}
\vspace{-3mm}
\caption{Two dimensional structure of cluster model in the meridional
         plane with $(W_0, \omega_0) = (6,0.6)$
         and with tidal boundary at $t/\tau_{rh,i} = 9.03$.
         One dimensional velocity dispersion (a) and rotational velocity (b)
         are shown.}
\label{fig-2d}
\end{figure}

The 2-dimensional structure of cluster with $(W_0, \omega_0) = (6,0.6)$ 
with tidal boundary in the meridional plane is shown in Fig. \ref{fig-2d}
at a time $t/\tau_{rh,i} = 9.03$.
Note that the axes are measured in units of initial core radius and that
the current core and half-mass radii have already decreased to 
$\sim 0.038 r_{c,i}$ and $\sim 0.64 r_{c,i}$, respectively. The one-dimensional velocity
dispersion shows a clear sign of collapsed core, i.e., linearly increasing
velocity dispersion toward cluster center. The morphology of two-dimensional
rotational velocity after core bounce shows a similar structure to that of
pre-collapse phase (see Paper I for comparison), although the degree of
rotation decreased.



\subsection{Velocity Dispersion and Angular Speed}

In Fig. \ref{fig-sigma}, we have shown the time evolution of 
central velocity dispersion ($\sigma_0$) and central rotational angular
speed ($\Omega_0$). As we have seen from the rotational curves in
Fig. \ref{fig-rotcv}, the central parts have the rigid body rotation,
and the angular speed is a constant out to large radius.

Both $\sigma_0$ and $\Omega_0$ increase with time rather slowly
until core collapse
and decrease afterward. The increase of $\Omega_0$ is a consequence of
the gravo-gyro instability. During the post-collapse, these quantities
drops rapidly with time. The general behavior of $\sigma_0$
does not seem to be affected by the presence of rotation. As we
will see in \S3.7, rotation energy is only a small fraction of the
total kinetic energy near the center throughout the evolution.

The relationship between the central density and $\sigma_0$, and $\Omega_0$ are
shown in Fig. \ref{fig-so}. For $\sigma_0$ versus $\rho_0$, we have
plotted all six models of Table 1, while only rotating four models are
shown for $\Omega_0$ versus $\rho_0$ figure. The velocity dispersion is 
nearly independent of initial rotation: all models with the same $W_0$ fall
on nearly single lines. 
The power-law behavior of $\sigma_0$ on $\rho_0$
during the pre collapse  phase is a consequence of self-similarity
of collapsing core. During this phase, it is well known
that $\sigma_0\propto \rho_0^{0.1}$ (e.g., Cohn 1980).
During the post-collapse phase, we can again apply the energy
balance argument to obtain $\sigma_0\propto \rho_0^{1/6}
M^{1/9}$. For isolated clusters, $M=constant$ and $\sigma_0
\propto \rho_0^{1/6}$. 
Since both $M$ and $\rho_0$ decrease with time, we expect that 
$\beta>1/6$ if we express $\sigma_0\propto \rho_0^\beta$.
Our result in Fig. \ref{fig-so} shows that $\beta \approx 0.23$.

The angular speed also appears to follow power law on
$\rho_0$ during the pre collapse phase, except for the early stage of
evolution for models with central potential of $W_0=3$.
Although the behavior of $\Omega_0$ during the post-collapse phase
depends on the amount of angular momentum loss by the end of the core-collapse,
it still has a power-law relation with central density, $\rho_0$.

\begin{figure}
\epsfig{figure=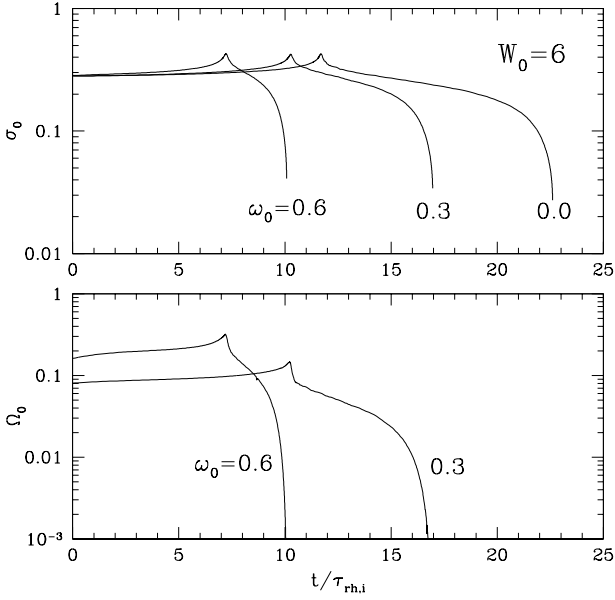, height=0.475\textwidth, width=0.475\textwidth}
\vspace{-3mm}
\caption{Time evolution of central velocity dispersion and central
angular speed for models with $W_0=6$ and $\omega_0=0.0$, 0.3 and 0.6.}
\label{fig-sigma}
\end{figure}

\begin{figure}
\epsfig{figure=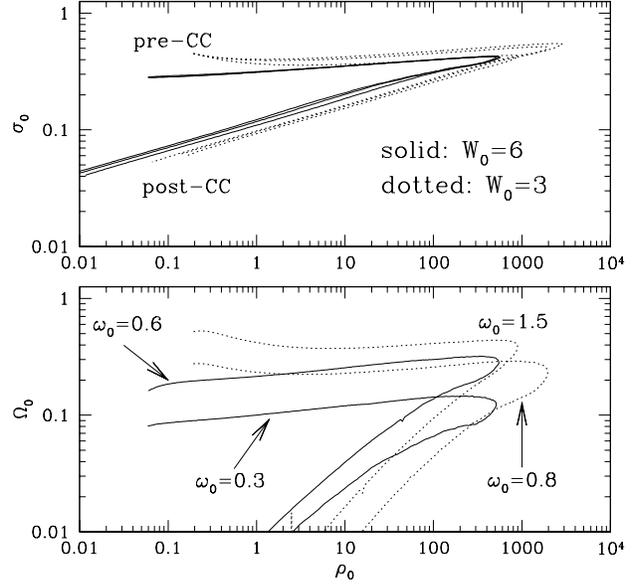, height=0.475\textwidth, width=0.475\textwidth}
\vspace{-3mm}
\caption{The evolution of $\sigma_0$ and $\Omega_0$ as a function of
$\rho_0$ for all six models shown in Table 1. The velocity dispersion 
follows power-laws during pre and post collapse phase, and is nearly
independent of rotation. The angular speed also follows near power-law
on $\rho_0$ during the collapsing phase.}
\label{fig-so}
\end{figure}

\begin{figure}
\epsfig{figure=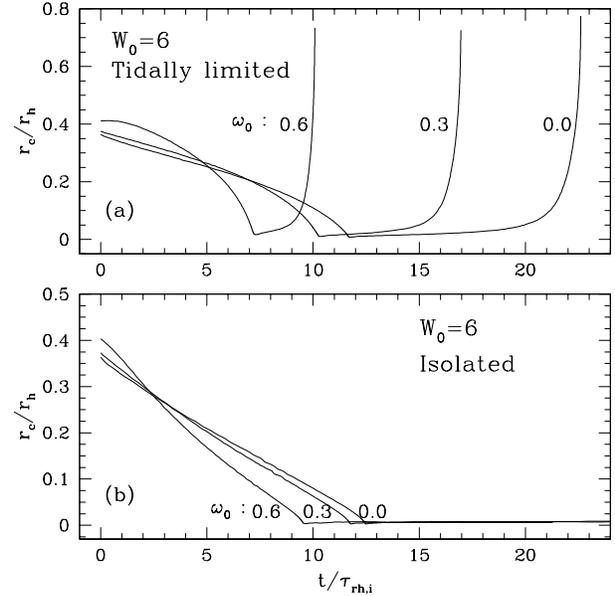, height=0.475\textwidth, width=0.475\textwidth}
\vspace{-3mm}
\caption{Ratio between core and half-mass radii for models with tidal
        boundary (a) and for models without tidal boundary (b). The central
        potential for both models is $W_0=6$.}
\label{fig-radii}
\end{figure}

\begin{figure}
\epsfig{figure=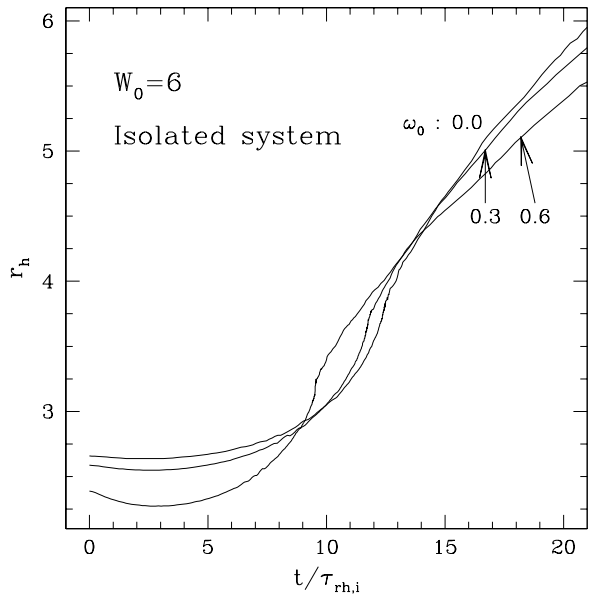, height=0.475\textwidth, width=0.475\textwidth}
\vspace{-3mm}
\caption{Comparison of half-mass radii for models with $W_0=6$, and without
         tidal boundary. The half-mass radius for model with
         higher value of initial rotation is seen clearly to be smaller 
         than the half-mass radii for models slower/no initial rotations.}
\label{fig-rh}
\end{figure}

\subsection{Core and Half-mass Radii}

The evolution of $r_c/r_h$ for models with central potential
of $W_0=6$ is shown in Fig. \ref{fig-radii}: the upper panel is for the 
tidally limited clusters and the lower panel is for isolated clusters.
For tidally limited clusters, more rapidly rotating initial models 
show larger value of $r_c/r_h$
due to smaller $r_h$ than slowly rotating models.
Both $r_c$ and $r_h$ decrease with time although there is a 
difference in decreasing rate depending on the initial degree of rotation.
For example, $r_c/r_h$ remains nearly constant in the early phase for our
$\omega_0 =0.6$ model, while other models show gradual decrease. 
After core bounce, both $r_c$ and $r_h$ increase, but
$r_c$ increases more rapidly than $r_h$.
In the late phase of the evolution $r_h$ decreases again due to 
a large amount of mass loss. The minimum values of $r_c/r_h$ are nearly
independent of $\omega_0$.

For isolated clusters, all models show similar value of 
$r_c/r_h \sim 0.007$ after core bounce. Again, this is due to
the fact that the cluster remains self-similar if the
post-collapse is dominated by three-body binary heating.
According to the self-similar model by Goodman (1987), $r_c/r_h \propto
N^{-2/3}$, and thus it should be factor of $\sim 1.84$ smaller than the
case with $N= 2000$: we find that our value is about a factor
of 2 smaller than the $N$-body calculation by Giersz \& Heggie (1994b)
with $N=2000$.

Since our system is axisymmetric,
it is a little bit difficult to define the half-mass radius. We used 
the same definition as in
Paper I. For early stages of the evolution, when the system is more flattened,
the method used here (and also in Paper I) may be problematic though
the difference is small. However, after core bounce we expect that the
difference between true half-mass radius and the half-mass radius computed
in the present study is negligible.

The evolution of $r_c/r_h$ depends sensitively on the choice
of boundary conditions.
Strong increase of $r_c/r_h$ after core bounce for tidally limited
clusters is mainly due to strong mass loss. The effect of rotation after
core bounce is shown in Fig. \ref{fig-rh}, which 
displays the evolution of
$r_h$ for isolated clusters with initial central potential of $W_0 = 6$.
After core bounce, $r_h$ for initially rapidly rotating cluster 
($\omega_0 = 0.6$) increases more slowly than 
the model without initial rotation. The difference of $r_h$
between two models is $\sim 6$ \% at $t/\tau_{rh,i} \sim 20$.
Although we tried to keep these
stellar systems isolated from any tidal effect,
there is some amount of mass loss through the computational boundary.
The amount of mass loss for our isolated
models is $\sim 2$ percent, irrespective of initial degrees of rotation at the
end of our calculation.
Therefore, we conclude that the difference between half-mass radii (in late
evolutionary stages) for different
initial rotation for isolated models is
due to rotation, which is still present near a radius of $r \sim r_h$. 
The smaller value of
$r_h$ for a more rapidly rotating cluster can be
explained by a transfer of $z$-component of angular momentum($J_z$) in outward
direction. So, in contrast to our findings (for the isolated cluster only)
that the post-collapse core structure, as represented by Cohn's dimensionless central
potential, does not depend on initial rotation, we find here that at the
half-mass radius the system still remembers about its different initial
degrees of rotation.

\begin{figure}
\epsfig{figure=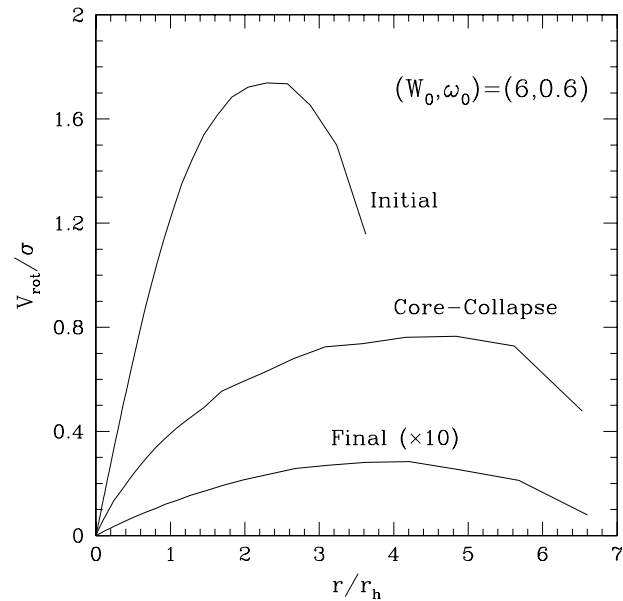, height=0.475\textwidth, width=0.475\textwidth}
\vspace{-3mm}
\caption{$V/\sigma$ as a function of radius in units of
half-mass radius for models with initial $W_0=6$,
and $\omega_0=0.6$ at three epochs as indicated in the figure. The 
location of maximum $V_{rot}/\sigma$ moves outward. }
\label{fig-vsigma}
\end{figure}

\subsection{$V_{rot}/\sigma$}

We try to give some evidence what could be the consequences of our
model simulations for the interpretation of velocities (rotational
and dispersion) in globular star clusters. It has to be emphasized,
however, that due to the idealized nature of our numerical studies
(e.g. equal point masses, no tides, no stellar evolution) such
discussion must be preliminary and can certainly not be done on
a quantitative, but rather on a more qualitative level.

$V_{rot}/\sigma$ has been analyzed in our model clusters, where $V_{rot}$ denotes the
rotational velocity, and $\sigma$ the 1D velocity dispersion, as a function of time and
radius. $V_{rot}/\sigma$ describes the relative importance
of rotational versus pressure support in the local cluster kinematics
and dynamics; while it has been very successfully used in
observational studies of elliptical galaxies and bulges of spiral galaxies
this quantity recently has become available also from globular cluster
observations (Gebhardt 2001, personal communication).

In Fig. \ref{fig-vsigma}, we have shown the runs of $V_{rot}/\sigma$
as a function of radius in units of half-mass radius for the model
with initial $W_0=6$ and $\omega_0=0.6$ at three epochs:
initial model, at core-collapse, and near final disintegration.
We notice from this figure that the position of the maximum $V_{rot}/\sigma$ 
relative to $r_h$ moves 
outwards with time. This is another indication of angular momentum
transport. Also note that $(V_{rot}/\sigma)_{\rm max}$ for the initial
model is much larger than the late values of an evolved cluster,
it decreases monotonically with time.
For this particular model, the $(V_{rot}/\sigma)_{\rm max}$ becomes 
around 0.8 at the core collapse, and further decreases in the
post-collapse phase to very small values smaller than 0.1.
Towards the centre the models presented in this paper all
show a decrease of $(V_{rot}/\sigma)$ to zero. This is consistent
with what we see in
Fig. \ref{fig-rotcv}, that the stars rotate like
solid body roughly inside half-mass radius, regardless of the
the evolutionary phase. Note that we obtained $V_{rot}$
and $\sigma$ in the rotational plane, and the observed
value of $V_{rot}/\sigma$ should be always be smaller than
this figure.

\begin{figure}
\epsfig{figure=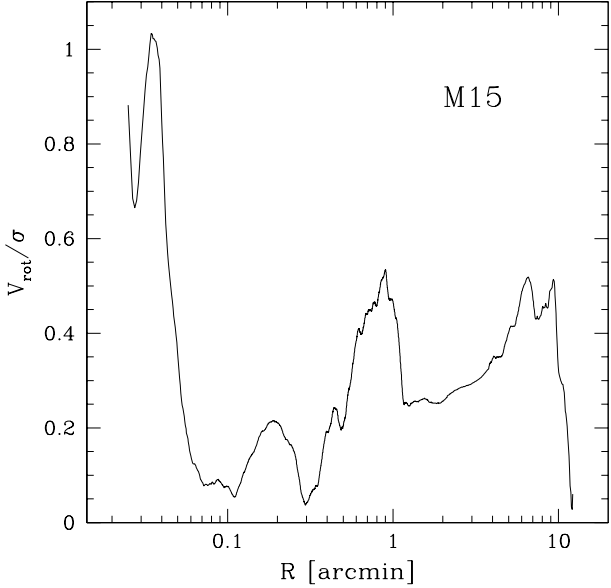, height=0.475\textwidth, width=0.475\textwidth}
\vspace{-3mm}
\caption{$V_{rot}/\sigma$ as a function of radius in arc minutes for M15 (data
provided by Gebhardt).}
\label{fig-gebhardt}
\end{figure}

Gebhardt (personal communication) provides some new data on 
direct measurements of
$V_{rot}/\sigma$ in M15, which are shown in Fig,~\ref{fig-gebhardt}. Regarding
the outer zones of the cluster, our
data are consistent with one of our post-collapse models starting at
$W_0=6$ and $\omega_0=0.6$, i.e., an initially fairly large amount of
rotation, as far as the core, half-mass and outer radii are considered.
But for the central regions he finds a strong rise in the observed
rotational speed, which is not present in our models.
Note, however, that multi-mass models obtained with an earlier version
of our code offer a possible explanation for this
phenomenon: the gravo-gyro instability effect (Hachisu 1979) is much more
pronounced if a mass spectrum is present as compared to the equal mass
model (see for that Fig.~4 of Paper 1). Heavy masses go into core bounce
with a strong acceleration of rotational velocity and a decrease of
dispersion velocity, so they would exhibit a high $V_{rot}/\sigma$ in the
centre (compare figure in Spurzem \& Einsel 1998, where the angular velocity
and the dispersion velocity of ten mass components in the centre are
shown as a function of time). We will further study this effect in
future work.

Though many of the Galactic globular clusters do not rotate significantly,
there have been rotation curves measured for some clusters earlier
(Meylan \& Mayor 1986; Lupton et al. 1987; Gebhardt et al 1994, 1995).
The clusters with measured rotation includes $\omega$ Cen, 47 Tuc, M13,
M15, and NGC6397. For most of these clusters, the
rotation measurements are done in a rather limited range and rotation
curves are hardly known. Also in earlier papers,
Gebhardt et al. (1995) made very
careful measurements of stellar velocity field using Fabry-Perot
spectrophotometer and found that the projected rotation velocity rises
or remains nearly flat from 30$^{\prime\prime}$ to 10$^{\prime\prime}$
for M15, 47 Tuc, and NGC 6397. As was discussed above this cannot
be reproduced in our equal mass models here yet, but we think
a multi-mass model could be able to explain the effect.
Also in the case of 47 Tuc the rotation curve seems to
be consistent with the solid body rotation up to $\sim 2^\prime$
corresponding to $\sim 0.7 r_h$,
although again the very central part appears to deviate from
the solid body rotation.

We need density and velocity dispersion profiles as well
as the rotation curve in order to estimate the relative rotational energy
(such as $-T_{rot}/E_{tot}$). Assuming same $V_{rot}/\sigma$ throughout the
cluster, we estimate $-T_{rot}/E_{tot} = 0.013\sim 0.05$.
The dynamical ellipticities expected lie in the range of 0.03 to 0.05,
depending on the inclination. The observed ellipticities of these
clusters are roughly consistent with the $V_{rot}/\sigma$, although
there are large uncertainties on the estimates of ellipticities.

 \section{Conclusion}

 The first self-consistent evolutionary models of rotating star
 clusters in the post core collapse phase have been obtained by
 improving an earlier version of a FOPAX 
 for axially
 symmetric systems. For that purpose the code used in previous models
 of pre-collapse rotating clusters (Einsel \& Spurzem 1999, Paper I)
 has been improved by including a three-body binary heating 
 and also in some of its numerical procedures.
 We start as initial models with generalized King models
 including rotation and study a few fiducial cases with
 dimensionless central potential $W_0$ 3 or 6, and
 a dimensionless rotation parameter $\omega_0$ from 0.0 (no rotation,
 spherical system) to 1.5 for $W_0=3$ and 0.0 to 0.6 for $W_0=6$.
 While most of the model star clusters include a tidal boundary modelled
 by an energy cutoff, we discuss one sequence of isolated models 
 ($W_0=6$) to distinguish the effects of internal relaxation and
 tidally induced mass loss.

 Our results show that, as
 in the pre-collapse case (compare with Paper I),
 the evolution of the tidally limited cluster is significantly accelerated 
 by initial rotation. Comparing with isolated models we find that in
 pre-collapse the acceleration of evolution is due to a combination of
 both internal relaxation (enhanced by the increased fraction of ordered
 motion in the system) and easier escape across the tidal boundary in the
 co-rotating direction. For post-collapse however, the isolated systems do
 not show any strong dependence of their structure and evolution due
 to initial rotation. However, in case a tidal field is present, the
 rotating cluster starts off into the post-collapse phase with a
 slightly larger ratio of core over half-mass radius $r_c/r_h$, and a
 slightly shallower value of the central potential, and this
 triggers a faster evolution during its complete lifetime, up to
 the final dissolution. The acceleration of its evolution is 
 comparable to that found in the pre-collapse state in Paper I.

Evidently, models with higher initial rotation dissolve faster in
the tidal field.
The effect of rotation disappears after $t/\tau_{rh,i} \sim 3$ for model
with initial rotation of $\omega_0=0.6$. The mass loss rate
after core collapse is different from each other depending on initial
degrees of rotation. However,
the main reason for the difference of
mass loss rate is a shallower scaled central potential for highly rotating model
at a time of core collapse
than the model without initial rotation. Therefore we conclude that
the overall shape of cluster changes very little after core collapse
irrespective of initial degrees of rotation.

The rotational energy decreases all the time as the cluster loses
mass, together with angular momentum. 
By the time of core collapse,
the rotational energy becomes some fraction of its initial
value. But still initially rotating and non-rotating clusters
exhibit different structure properties at intermediate radii,
e.g. in their ratio $r_c/r_h$, and the $(V_{rot}/\sigma)$ value
(rotational velocity over 1D velocity dispersion) has a clear
maximum as a function of radius, and the radial position of this
maximum (measured in units of the half-mass radius) goes out. The
maximum value of $(V_{rot}/\sigma)$ at a given time decreases monotonically
with the dynamical age of the cluster, even across core collapse.
We stress that this is a possible unambiguous dynamical clock
to measure dynamical ages of clusters, which is not reset to zero
as the cluster undergoes core bounce and re-expansion. Its present
value allows also to put certain limits on the initial amount
of rotational energy in clusters. Our values of $(V_{rot}/\sigma)$
can be confronted with much improved recent observations of
rotation curves in globular clusters (Gebhardt 2001, personal communication).
We can pinpoint a post-collapse evolutionary state where our model
cluster has a maximum value $(V_{rot}/\sigma)_{\rm max}$ which matches the
observed value, e.g., in M15; on the other hand our present models do
not reproduce the observed strong increase in central rotation, we rather
have always a rigid body rotation in the central regions; but
preliminary multi-mass models (Spurzem \& Einsel 1998) are indeed
able to explain the observed effect. It is a strongly 
pronounced occurrence of what was found by Hachisu (1979) as 
gravo-gyro catastrophe; it is much less
pronounced in the equal mass system, but the heavy masses in a
multi-component system are strongly subject to it (Spurzem \& Einsel 1998).

Most of the results discussed in this paper, however, are still derived
from an equal mass model.
Actual clusters should have a mass spectrum, and the observed rotation
curves are usually dominated by the brightest component. In order
to compare with the observed clusters, we need to further study the
multi-mass models rather than equal-mass models.
Also, we treat the cluster
as a 2D model in phase space only (distribution function $f=f(E,J_z)$ only
depends on two constants of motion, energy, and $z$ component of
angular momentum). So all relaxation effects (diffusion coefficients)
do not discriminate between different orbits having same $E$ and $J_z$,
but different third integral $I_3$. In Paper I some discussion of
the possible errors is given by measuring the artificial flattening
of an initially spherical system treated by our axisymmetric model.
Here we do not continue this, but propose for the near future an
extensive study of direct $N$-body models of rotating clusters, similar in scope
to the seminal results obtained for non-rotating clusters by Giersz \& Heggie
(1994a, 1994b, 1996). This will make it possible to assess the validity
of our approximations. Preliminary studies of a few of such models
do exist (Boily 2000, Boily \& Spurzem 2000) and they support the
Fokker-Planck results, but they do by any means not provide a broad
enough statistical and parameter range.

\section*{Acknowledgments}
We are grateful to generous support from BK21 program. This work
was partially supported by the interdisciplinary grant from
Seoul National University. R.Sp. wants to thank the Dept. of
Astronomy, Seoul National Univ., and in particular H.M. Lee,
E. Kim, and many other colleagues there for generous and very friendly 
hospitality during a research visit in Korea. 
Part of this work has been completed
during a research stay at the Univ. of Tokyo under the collaborative
German-Japanese research grant DFG/JSPS 446 JAP-113/18/0-2. R.Sp.
thanks the colleagues at Univ. of Tokyo for friendly and supportive
hospitality.
E.K. \& M.G.L. are supported in part by the MOST/KISTEP 
international collaboration
research (1-99-009).

 \end{document}